\documentclass[aps,prb,onecolumn]{revtex4-1}
\usepackage{latexsym}
\usepackage{graphicx}
\usepackage{dcolumn}
\usepackage{bm}
\usepackage{amssymb}
\usepackage{amsmath}
\usepackage[space]{grffile}
\usepackage{color}

\begin{document}

\title{Model Predictions for Time-Resolved Transport Measurements Made near the Superfluid Critical Points of Cold Atoms and $K_3 C_{60}$ Films}
\date{\today}
\begin{abstract}
Recent advances in ultrafast measurement in cold atoms, as well as pump-probe spectroscopy of $K_3C_{60}$
films, have opened the possibility of rapidly quenching systems of interacting fermions to, and across, a
finite temperature superfluid transition. However, determining that a transient state has approached a
second-order critical point is difficult, as standard equilibrium techniques are inapplicable. We show that
the approach to the superfluid critical point in a transient state may be detected via time-resolved transport
measurements, such as the optical conductivity. We leverage the fact that quenching to the vicinity of the
critical point produces a highly time dependent density of superfluid fluctuations, which affect the
conductivity in two ways. First, by inelastic scattering between the fermions and the fluctuations, and
second by direct conduction through the fluctuations, with the latter providing a lower resistance current
carrying channel. The competition between these two effects leads to nonmonotonic behavior in the 
time-resolved optical conductivity, providing a signature of the critical transient state.
\end{abstract}
\author{Yonah Lemonik\textsuperscript{*}}
\author{Aditi Mitra\textsuperscript{*}}

\affiliation{Department of Physics, New York University, 726 Broadway, New York, NY, 10003, USA}

\maketitle

Measurement techniques using ultra-fast optics have enabled the study of fermionic fluids on time scales far shorter than the timescale of
thermalization~\cite{Fausti11,Smallwood14,Averitt16}.
Similarly, in ultra-cold atomic systems the relaxation times are long enough that pre-thermalization behavior may be studied~\cite{Gring12,Langen15,Bloch15}.
In both cases experimental techniques presently exist to study the collective behavior of fermions in the period before they have relaxed to their equilibrium
or steady state behavior. As an example, a transient
state with superconducting-like optical properties was produced by ultra-fast laser stimulation of K$_3$C$_{60}$
films~\cite{Mitrano15}.

However there is a difficulty in determining whether a transient state is in any sense related to any particular phase.
More precisely one may ask whether a non-equilibrium
system has been taken through the vicinity of a certain second-order phase transition. For example, whether a state with "superconducting-like"
optical properties is in fact related to the equilibrium superconducting phase transition, or whether it is some other transient state.
Equilibrium methods for detecting such phase transitions such as the specific heat are not applicable to ultra-fast or non-equilibrium settings.

We suggest that this difficulty may be resolved by looking for signatures of time-dependent fluctuations in the experimental data.
Fluctuations increase in a singular fashion in the vicinity of a second-order phase transition, and therefore may be used as a signature
that such a transition has been approached.
In particular we give predictions for time-resolved transport measurements of fermions
quenched close to the superfluid critical point. In the process we generalize theoretical treatments for
equilibrium critical systems to the strongly non-equilibrium regime.

\begin{figure}
\includegraphics{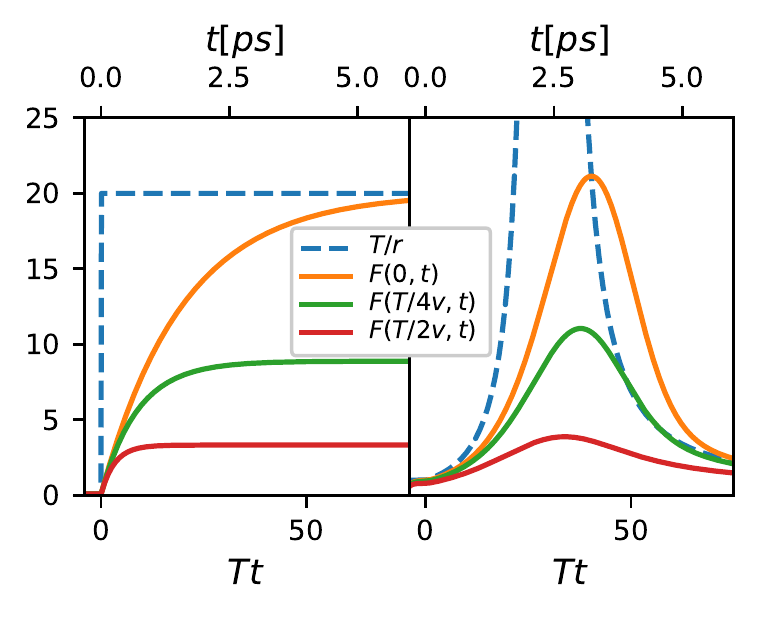}
\caption{Growth of superfluid fluctuations $F(q,t)$ following an interaction quench.
Fluctuations at several different momenta $q$ are shown. The times are
measured in units of $T^{-1}$ (lower axis) and in terms of ps (upper axis) for $T\sim 100$K.
The quantity $T/r$ is the inverse of the detuning from the critical point,
which at equilibrium is equal to $F(q =0)$. Left: hard quench from the normal state to $T/r = 20$.
Right: Soft quench, with $r(t>0) = T[1-(t/t_*)e\exp(-t/t_*)]$, $t_*T = 30$.  \label{fig:Flucs}}
\end{figure}

The experimental setup proposed is a gas of fermions, initially at equilibrium at finite temperature.
This may be a gas of cold atoms in an optical lattice, or electrons in a thin solid state film. An attractive interaction is
turned on at a certain rate.
In the case of cold atoms this may be achieved by the tuning of optical resonances~\cite{Bloch08}. For a thin film, it may be achieved by
strong optical pumping of a phonon mode~\cite{Knap16,Kennes17,Sentef17,Sentef17b,Werner17}.
An attractive interaction causes superfluid fluctuations to develop,
leading to two possibilities. If these fluctuations are sufficiently strong, then a superfluid order
will spontaneously develop, as happens in equilibrium beyond the critical interaction strength.
The second possibility is that superfluid fluctuations are enhanced, see Fig.~\ref{fig:Flucs}, but spontaneous long range order does not develop.
Note that the interaction strength may be instantaneously supercritical, but the system remains disordered ({\sl i.e.}, no long range order) if
there is not sufficient time for the fluctuations to develop. In this paper we consider such disordered regimes.

In order to describe the fluctuation physics we must go beyond time-dependent mean field methods~\cite{Foster17, Kennesprb17}. We  develop a quantum kinetic equation,
derived within the two-particle
irreducible (2PI)
framework~\cite{CornwallJackiwTomboulis}, which describes the joint evolution of the unbound fermions and the superfluid fluctuations.
These equations predict the time-resolved transport properties of the interacting system. The theory is
formally controlled by a parameter $1/N$ which represents the size of the fluctuations. Our predictions
depend on only a small number of parameters, and do not depend on the precise form of the
fermion-fermion interaction or on the details of the band structure. Our calculation is valid for a clean system
and only if the fluctuations are not very strong, as given by a non-equilibrium equivalent of the Ginzburg-Levanyuk criterion \cite{Larkin00} specified below.

The kinetic equations incorporate two scattering processes. The first is incoherent
Andreev reflection of fermions off superfluid fluctuations. This leads to an
aging effect where the time-resolved optical conductivity decreases after the quench with a
characteristic power law. The second scattering process is the binding and unbinding of fermions into Cooper pairs,
leading to the growth in fluctuations after the quench, and to
a non-equilibrium Azlamazov-Larkin-like effect~\cite{Larkin00}, where the Cooper-pair fluctuations
serve as an additional low resistance channel for the current.

\begin{figure}
\includegraphics[width = \columnwidth]{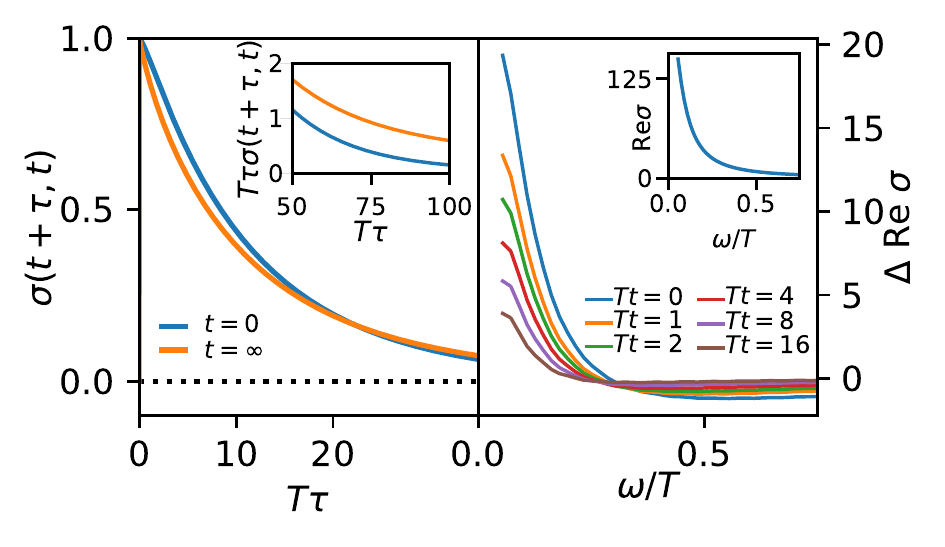}
\caption{Conductivity for the hard quench. Left: Conductivity $\sigma(t + \tau, t)$ as a function of $T\tau$
for $t = 0$ (blue) and $t\rightarrow \infty$ (orange).
The parameters are  $T \tau_r = 5$, $\alpha = 0.5$ and detuning $r=0$.
In the inset are the tails of $\sigma(t+\tau,t)$, $T \tau> 30$, plotted as $T\tau\sigma(t +\tau,t)$ to improve visibility.
The curves asymptote to a constant as explained in the main text. 
Right: Re $\left[\sigma(\omega,t) - \sigma(\omega, t \rightarrow\infty)\right](e^2/\hbar)$
for different times since the quench. Note the times increase in a geometric fashion.
Inset: Re$\sigma(\omega, t\rightarrow\infty)$. This diverges as $\log\omega$ as $\omega \rightarrow 0$ so all curves are clipped at $\omega = .01T$
\label{fig:sumfig}}
\end{figure}

The kinetic equations describe the conductivity for arbitrary dependence of the interaction on the time.
We give results for two different quench protocols, "hard" and "soft".
In the hard quench (see left panel Fig.~\ref{fig:Flucs}) the interaction is instantaneously changed to be in the vicinity of the equilibrium
phase transition. We show that there are power-law corrections
to the conductivity at times close to the quench.

We also discuss a soft quench where the interaction is smoothly switched on and off (see right panel Fig.~\ref{fig:Flucs}), which is more applicable
to K$_3$C$_{60}$ films where the transient state survives for $2$-$10$ps~\cite{Mitrano15}.
We find that the optical conductivity  evolves non-monotonically in frequency, providing
a strong signal for time resolved fluctuation effects and the existence 
of a transient lower resistance current carrying channel.

\subsection{Model and 2PI Formalism}
We consider a model with spinful fermions in a potential (optical or crystal) lattice, without disorder.
The electrons interact via an on-site attractive interaction $U(t)$, which is allowed to vary with time. In equilibrium,
at temperature $T$, the critical interaction strength $U_c(T)$ separates the disordered and superfluid phases.

We employ the $2$PI~\cite{Ivanov98} formalism, which generates equations of motion for two-operator correlations, such as the
fermion Green's functions. The formalism takes a single generating
functional $\Gamma\left[\cdot \right]$, which we take corresponding to the random phase approximation (RPA)
justified by a large-N approximation~\cite{Lemonik17}.
The resulting equations are~\cite{Suppmat}
\begin{equation}
g^{-1}\!\circ G = \frac{i}{N}\!\left(DG \right)\circ G;
\quad D^{-1}[G] \equiv \!U^{-1}\!(t)-\Pi[G],
%\Pi[G] \equiv iG\cdot G .
\label{eq:quantumkinetic}
\end{equation}
where $\Pi[G] \equiv iGG$, $\circ$ implies convolution, $g$ and $G$ are the non-interacting and interacting Green's functions respectively,
while $D$ is the propagator of the superfluid fluctuations, with
the equal time Keldysh~\cite{Kamenevbook2011} component representing the size of the fluctuations $F(q,t)$ (see Fig.~\ref{fig:Flucs}).

The functionals $D,\Pi$ depend self-consistently on $G$,
thus Eq.~\eqref{eq:quantumkinetic} represents a highly non-linear set of equations.
We consider an initial state of free fermions at a non-zero temperature $T$.

\subsection{Fluctuations}
In an equilibrium system of fermions, with an attractive interaction
below the critical interaction, the expectation value $\langle \Delta(\vec{q} = 0)\rangle$ vanishes, where
$\Delta^\dagger (\vec{q})\equiv \sum_k c^\dagger_{k}c^\dagger_{q-k}$ is the operator that creates a Cooper pair of momentum $q$, and $c^\dagger_k$
creates a single fermion. However, superfluid fluctuations $F_{\rm eq}(q)$
defined by $F_{\rm eq}(q) \propto \langle |\Delta^\dagger(\vec{q})|^2 \rangle$ increase as the critical point is approached (setting $k_B=\hbar=1$),
\begin{equation}
F_{\rm eq}(q)  = \frac{T}{v^2|q|^2/T + r}; \quad v q \ll T,\quad r\ll T,\label{Fss}
\end{equation}
$v$ being the average Fermi velocity, and the detuning $r \propto U_c(T) - U$.
Thus the long wavelength fluctuations become pronounced as $r\rightarrow 0$.

The non-equilibrium dynamics are governed by two energy scales:
the temperature $T$ and the detuning from the
critical point $r\propto U- U_c(T)$. However, for a system close to the critical point with $T \gg r$ the dynamics
are essentially classical, as quantum fluctuations are averaged out on the time scale $T^{-1}$, but the dynamics occur on the scale $r$.
Thus the quantum kinetic equation can be reduced to a
joint kinetic equation for the evolution of the occupation numbers of the fermions and the distribution of fluctuations~\cite{Suppmat}.

The dynamics for the fluctuations are,
\begin{equation}
\left(\partial_t + r(t) + \frac{v^2|q|^2}{T}\right)F(q,t) = T,\label{eq:Fode}
\end{equation}
We emphasize that $F(q,t)$ is not given by the instantaneous value of $r(t)$, but rather by the full history of $r(t)$. This yields non-trivial dynamics.
Note that if $v q \gtrsim T$, then the $F(q,t)$ will equilibrate
on the short time $1/T$. Thus the non-trivial time dependence is concentrated in the modes $q v \ll T$ and it is sufficient to work with them.
Eq.~\eqref{eq:Fode} is valid only when $F(q=0,t) \ll E_F/T$~\cite{Suppmat}. This reduces to the Ginzburg-Levanyuk criterion $r \gg T^2/E_F$ in equilibrium.

\subsection{Conductivity}

We now discuss methods to detect the aging of superfluid fluctuations. A signature in photo-emission discussed
elsewhere~\cite{Lemonik17}, showed that growing fluctuations lead to a decreasing fermion lifetime, given by a universal scaling form.
Here we discuss the signatures in time-resolved transport experiments. The transport is studied by
varying the 2PI action, or the quantum kinetic equations, in response to an external electric field~\cite{Suppmat}.

We begin from the definition of the conductivity
$J(t) = \int dt' \sigma(t,t') E(t')$,
where all quantities are constant in space.
There is no single notion of a Fourier transform in a non time-translation invariant setting, but for the purposes of demonstration,
we study the behavior of the transform ${\sigma}(\omega, t) = \int\! d\tau\, \sigma(t + \tau,t)\exp\left(-i\omega \tau\right)$.
Note that $\sigma(\omega,t)$ depends on the state at times $>t$, especially at small $\omega$. In order to quantify the departure
from simple ohmic behavior, we plot the frequency dependent relaxation time
$\tau_{\rm Dr}(\omega) =-\text{Im}\left[\sigma\right]/ \left[\omega\text{Re}\sigma\right]$.
This is a frequency independent constant when the conductivity is governed by the Drude law $\partial_tJ(t) = -\tau^{-1}_{\rm Dr} J(t)$.

In a fluid without a lattice potential or disorder, Galilean invariance and conservation of momentum forces $\sigma(t,t') \propto \theta(t-t')$,
so that the current never relaxes. However the combination of a lattice potential and interaction causes a fraction of the current to relax,
even without disorder or Umpklapp scattering. The dynamics of the momentum decouple from the relaxing current $J$~\cite{Suppmat}
which obeys the following equation in $d$ spatial dimensions~\cite{Suppmat},
\begin{align}
&\partial_t J(t) -\frac{\rho}{\bar{m}} E(t) = -\frac{1}{\tau_r} \biggl(A(t) J(t) \nonumber\\  &\quad-\alpha \int_0^t dt'\biggl[B(t,t') J(t')
+ \frac{\rho}{2\bar{m}} C(t,t') E(t') \biggr]\biggr),\label{Jkin}\\
&A(t) \equiv T^{-\frac{d}{2}}\!\int_0^t\!ds\, \frac{e^{-\int_s^t \!du\, r(u)}}{(t-s)^{d/2+1} },\,\,
C(t,s) \equiv\! \int_0^s\!ds' B(t,s') ,\label{eq:Adef}\\
&B(t,s)\! \equiv T^{-\frac{d}{2}}\! \int_0^s\!dt'\frac{\left(1+\frac{d}{2}\right) + r(t)(t-t')}{(t-t')^{d/2+2} }e^{-\int_{t'}^{t}\!du\, r(u)}.
\end{align}
Above $\rho$ is the density of fermions, $\bar{m}$ is the average effective mass,
and $\alpha,\tau_r$ are material dependent parameters related to the extent to
which Galilean invariance is broken. Without loss of generality we set $\rho/{\bar m}=1$.

These equations are the central result of this paper. They may be solved numerically for an arbitrary $r(t)$ and electric field,
giving the conductivity $\sigma(t,t')$. We spend the remainder of the paper analyzing the implications of these equations for $d=2$.

The equation for the relaxational current $J$ in Eq.~\eqref{Jkin}
has two components. The $A(t)$ term represents incoherent small angle scattering of fermions
off of  superfluid fluctuations. The second term, proportional to $\alpha$, is a "memory" term, which is conceptually similar to the Azlamazov-Larkin term studied in
the theory of equilibrium fluctuation superconductivity. It can be understood as coming from the binding of fermions into
long-lived fluctuating Cooper pairs,
which decay at time $\gg T^{-1}$. This term is non-local in time because the dynamics of the fluctuations are governed by the time scale
$r^{-1}$, which diverges at the critical point.

An electric field  at time $t'$ directly
affects the  fluctuations altering the
distribution of the Cooper pairs, which then affect the current at
a much later time $t$. This is the origin of term $C$, which only appears because the fluctuations are charged. Thus the kinetic equation is quite different than for other kinds of critical points, such
as magnetic critical points, where the order parameter is neutral.

\subsubsection{Hard quench and aging}

\begin{figure}
\includegraphics[width =\columnwidth]{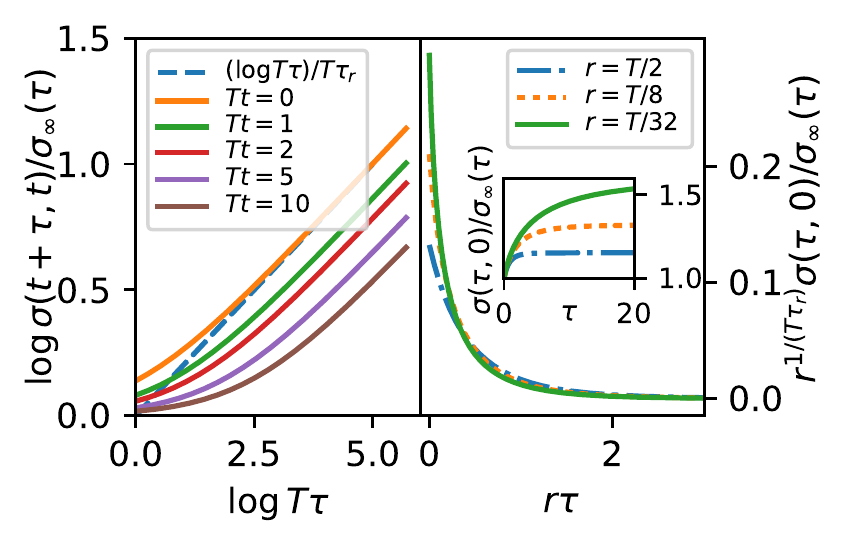}
\caption{
Aging in the limit $\alpha = 0$ for the hard quench.  Left: Log-Log plot of the  ratio $\sigma(t+\tau,t)$ and $\lim_{t\rightarrow\infty}\sigma(t+\tau,t)$ against $T\tau$.
The conductivity is enhanced at early times because of the absence of superfluid fluctuations.  In the case of $t =0$ the ratio converges to
$\left(T\tau\right)^\gamma$, where $\gamma = 1/T\tau_r$.
Right: Scaling plot of the $\sigma(\tau,0)$ at $\alpha =0$ for different detunings $r$. Inset: unscaled $\sigma(\tau,0)$.
Plots shown for $T\tau_r = 5$. \label{fig:scaling}
}
\end{figure}

We now seek signatures of criticality in the transient regime, first considering the hard
quench where $r(t) \gtrsim T$ for $t<0$ and $r(t) = r$ for $t>0$. We begin by studying the fully critical case,
$r=0$, where the functions $A$,$B$,$C$, simplify to
\begin{align}
&A(t) = 1- (T t)^{-1},\,\, B(t,t') = (Tt - Tt')^{-2} - (Tt)^{-2},\nonumber\\
&C(t,t') = (Tt-Tt')^{-1}-T(t+t')(T t)^{-2}.
\end{align}
The conductivity $\sigma(t+\tau,t)$ for these parameters is shown in Fig.~\ref{fig:sumfig}. The slow relaxation leads to "aging" phenomena,
where the conductivity approaches its equilibrium value not on the microscopic time scale $T^{-1}$, but much slower, as $r^{-1}$.
In particular at $r = 0$ the approach is via scale free power laws. There are two aging effects, one arising from the
dissipative term $A(t)$ which affects the short time conductivity, and the other due to the Cooper channel $C(t)$ which affects the
long time conductivity.
The effect of the latter is shown in
Fig.~\ref{fig:sumfig} (left, inset), where for $t\rightarrow\infty$ (orange line), 
the conductivity decays with a long tail ($\sim \tau^{-1}$).
However the tail is absent for $t= 0$ (blue line) as there are no fluctuations at $t=0$.

The short time aging is qualitatively different. It arises because $A(t)$ converges to its equilibrium value as $1/t$, causing
the conductivity at short times after the quench to be larger
than its equilibrium value. In particular, at short times, setting $\alpha = 0$ in Eq.~\eqref{Jkin} gives,
\begin{equation}
\sigma\!\left(t+ \tau,t\right) \rightarrow_{Tt\gg 1} e^{- \tau/\tau_r};\,\,
 \sigma\!\left(\tau,0\right)  =  \tau^{1/T\tau_r} e^{- \tau/\tau_r}.
\end{equation}
This behavior is shown in Fig.~\ref{fig:scaling}.
The strong power law amplification, with material dependent exponent,  $(T\tau_r)^{-1}$, is a consequence of the scale free nature of
the $r=0$ critical point, and thus provides an example of critical aging.
Since the $t=0$ conductivity is enhanced at short times by the lack of inelastic scattering, but is suppressed due to
the lack of fluctuation conductivity at long times, the aging is curiously non-monotonic in time as seen in Fig.~\ref{fig:sumfig} (left panel).
In Fourier space, Fig.~\ref{fig:sumfig} (right panel), aging manifests as a nearly logarithmic approach to the long time behavior of {\text Re}$\sigma$ at low frequencies.
Overall the dominant effect is a spreading of the initially very sharp peak in the initial state (due to weak scattering) to the wider line-shape in the inset. The conductivity is not Drude-like: as a consequence of the $\tau^{-1}$ tail of $\sigma(t+\tau,t)$, {\text Re}$\sigma(\omega,t\rightarrow\infty)$ diverges as $\log \omega$, as $\omega\rightarrow 0$.

We now consider the effect of finite detuning $r$ from the critical point.
The effect of the detuning is to suppress the aging behavior.
This may be seen analytically in the $\alpha = 0$ limit, where we find that for $r t\gg 1$,
$\sigma\left(\tau,0\right) \rightarrow (T/r)^{1/T\tau_r}\exp(-\tau/\tau_r)$, and the power law amplification saturates to a constant.

Similar to an equilibrium phase transition, the fact that the dynamics is controlled by the single diverging scale $r$ means that the
functions $B(t,t')$ and $C(t,t')$ can be expressed as a function of the quantity $r t$,$rt'$, and the function $A(t)$ may be written as $A(t) = c +  \frac{r}{T} g(r t) $, for some order one constant $c$.
As a result of this scaling behavior, for $\alpha \ll 1$, the ratio of short and long time conductivity is "universal",
\begin{equation}
\sigma(\tau,0)/\lim_{t\rightarrow\infty}\sigma(t+\tau,t) = r^{-1/(T\tau_r)}F_s(r\tau),
\end{equation}
for some scaling function $F_s(x)$. This scaling is demonstrated in Fig.~\ref{fig:scaling}, right panel.

\subsubsection{Smooth quench protocols}

\begin{figure}
\includegraphics[width = \columnwidth]{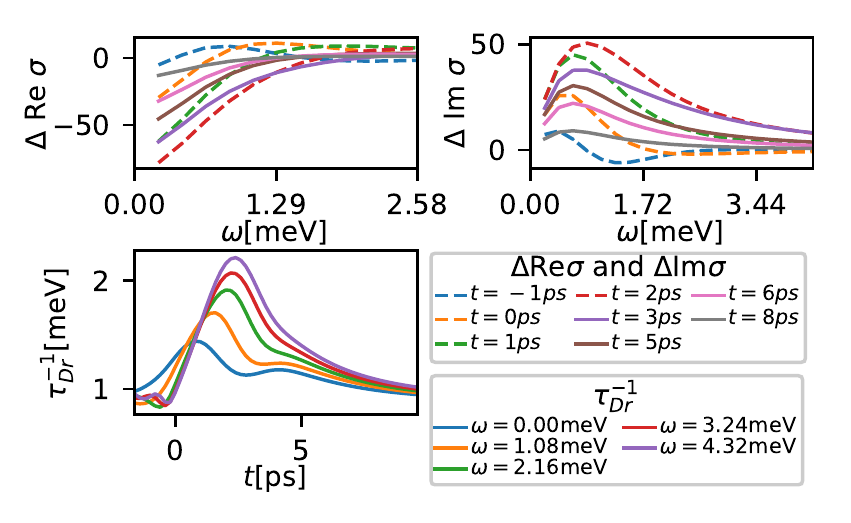}
\caption{Optical conductivity (in $e^2/\hbar$) 
for the soft quench with the profile for the detuning $r(t)$ shown in right panel of Fig.~\ref{fig:Flucs}.
Top panel: Re,Im $\left[\sigma(\omega,t)-\sigma(\omega,t=\infty)\right]=\Delta\sigma$
for several values of $t$ and $T=100$K. Early times $t\leq 2$ps are shown with dashed lines, later times $t>2$ps in full lines. Lower panel:
Drude parameter as a function of time for several frequencies. \label{fig:td_quench}}
\end{figure}

The hard quench may not accurately describe the behavior of pumped K$_3$C$_{60}$ films.
In these experiments, the effective fermion interactions may vary smoothly on the scale $T^{-1}$, as energy is slowly transferred from
the laser to the active phonon mode, and then to the environment.

Thus we consider a "soft" quench where $r$ smoothly decreases to zero and then smoothly increases back to a large value,
sketched in the right panel of Fig.~\ref{fig:Flucs} for the particular choice
$r(t>0)/T = 1-(t/t_*)e\exp(-t/t_*)$ where $Tt_* = 30$. We consider the regime where $t_* \gg \tau_r \gg T^{-1}$.
The time dependent optical conductivity is shown in Fig.~\ref{fig:td_quench} for $T=100$K for which $t_*=2.5$ps.
The primary effect seen in the top left panel is a strong dip in Re$\Delta\sigma$ around
$t \sim t_*$ for low frequencies $\omega \ll T$, but a peak at $t\sim t_*$ for higher frequencies $\omega \sim T$.
This can be understood as a spreading of the peak of Re$\sigma$, as in the hard quench. This behavior is accompanied by a peak in
Im$\Delta\sigma$ (upper right panel) at low frequencies at $t\sim t_*$.

Fig.~\ref{fig:td_quench} lower panel shows $\tau_{\text{Dr}}(t,\omega)$.
This parameter varies strongly with frequency during the quench.
This highly non-Drude behavior may be understood as coming from the two different effects that increasing fluctuations have 
on the conductivity.
The inelastic scattering effect - the $A$ term - leads to increasing scattering as a function of time, at high frequencies.
The memory effect - the $B$ and $C$ terms - leads to decreasing scattering as a function of time, at low frequencies.
As these are sensitive in different ways to the trajectory of
$r(t)$, the total effect is non-monotonic in time. Further the peak in $\tau_{\text{Dr}}(t,\omega)$
happens at different times for different frequencies. These non-monotonic behaviors and the dispersion in frequency are a consequence of
slow charged modes, and therefore a necessary consequence of proximity
to a second order superfluid transition.

\subsubsection{Conclusions}

We have discussed the effect of non-equilibrium superfluid fluctuations on the conductivity of fermions in a lattice potential,
 with applications to cold atoms and
pump-probe experiments. Two primary effects were discussed, a characteristic aging of the conductivity at short times,
and a power law tail of the conductivity at long times. The variation of these two effects
with detuning from the critical point, and corresponding signals in frequency space were discussed.
A key feature of crossing a superfluid critical point is a non-monotonic frequency dependence
of the optical conductivity.

Our results generalize to quenches below $T_c$ provided the transient state is too
short-lived for long range order to develop. However future experiments which manage to create transient states of a 
longer duration require extending the study of optical conductivity to include coarsening, followed by vortex dynamics, and 
eventually the role of strong phase fluctuations.
This work provides a theoretical treatment missing in the current literature on how to analyze transport in transient superfluids, and 
may be generalized to include disorder~\cite{Lemonik18a}, and explore hydrodynamic features in other transport properties.

{\sl Acknowledgements:}
The authors thank Andrea Cavalleri for valuable discussions. 
This work was supported by the US National Science Foundation Grant NSF-DMR 1607059.

%\bibliography{quench}


\newpage

\begin{center}  \Large\bf Supplemental Material \end{center}
\vspace{0.5cm}
\noindent\hspace{3cm} I.\quad Model and Quantum Kinetic Equation \dotfill 9\hspace{4cm}

\noindent\hspace{4cm} A.\quad Conservation Laws \dotfill 10\hspace{4cm}

\vspace{0.2cm}

\noindent\hspace{3cm} II.\quad Reduction to a quasiclassical kinetic equation \dotfill 11\hspace{4cm}

\noindent\hspace{4cm} A.\quad Linear response to an Electric field \dotfill 12\hspace{4cm}

\noindent\hspace{4cm} B.\quad Single mode projection \dotfill 13\hspace{4cm}

\vspace{0.2cm}

\noindent\hspace{3cm} III.\quad Evaluation of kinetic coefficients \dotfill 14\hspace{4cm}

\vspace{0.2cm}

\noindent\hspace{3cm} IV.\quad Time dependent detuning \dotfill 19\hspace{4cm}

\noindent\hspace{4cm} A.\quad Steady state behavior \dotfill 20\hspace{4cm}

\vspace{1cm}\section{Model and Quantum Kinetic Equation}\label{model}
We study a quench where the initial Hamiltonian is that of free fermions coupled to a translationally invariant, but time-dependent electric field $E(t)$,
\begin{eqnarray}
&&H_i=\sum_{k,\sigma=\uparrow,\downarrow,\tau=1\ldots N}\epsilon\left(k + A(t) \right) c_{k\sigma \tau}^{\dagger}c_{k\sigma \tau}.
\end{eqnarray}
Above $k$ is the momentum, $\partial_t A(t) = - E(t)$, $\sigma=\uparrow,\downarrow$ denotes the spin, and $\tau$ is an orbital quantum
number that takes $N$ values.
We consider the initial state to be the ground state of $H_i$ at non-zero temperature $T$, and chemical potential $\mu$. For simplicity we discuss the hard quench first. The generalization to arbitrary quench protocols will be outlined in Section~\ref{tdq}.

The time-evolution from $t>0$ is in the presence of a weak pairing interaction $U$. We write the quartic pairing interaction
in terms of pair operators $\Delta_q$ such that,
\begin{eqnarray}
&&H_f = H_i + \frac{U}{N} \sum_{q}\Delta^\dagger_{q}\Delta_{q},\nonumber \\
&&\Delta_{q} = \sum_{k\tau} c_{k,\uparrow,\tau}c_{-k+q,\downarrow,\tau};
\Delta^\dagger_{q} = \sum_{k,\tau} c^{\dagger}_{-k+q,\downarrow,\tau}c_{k\uparrow \tau}^{\dagger}.\label{Hf}
\end{eqnarray}
The Hamiltonian above assumes contact interaction, so that only fermions
with opposite spin quantum numbers scatter off of each other. The fermionic Green's functions are,
\begin{align}
G_R(1,2)&=-i\theta(t_1-t_2)\langle \{c(1), c^{\dagger}(2)\}\rangle,\nonumber\\
G_K(1,2)&=-i\langle [c(1), c^{\dagger}(2)]\rangle.
\end{align}

We here and generally suppress the spin and orbital indices, use numbers to indicate space-time coordinates, and do not define the advanced function
since it is the hermitian conjugate of the retarded part: $G_A(1,2) = G_R^*(2,1)$.
In the non-interacting case these are given by
\begin{align}
g_R^{-1} &\equiv i\partial_t - \epsilon(k + A(t)) \Rightarrow g_R(k,t_1,t_2)=-i\theta(t_1-t_2)e^{-i\int_{t_2}^{t_1}dt' \epsilon(k+A(t'))},\\
g_K(1,2) &\equiv \biggl[g_R(1,2) - g_A(1,2)\biggr]n(k,{\rm min}(t_1,t_2)) \Rightarrow g_K(k,t_1,t_2)= -ie^{-i\int_{t_2}^{t_1}dt' \epsilon(k+A(t'))}n(k+A({\rm min}(t_1,t_2))).
\end{align}

In what follows we employ the two particle irreducible ($2$PI) formulation to treat this problem. In the $2$PI approach the Green's functions $G$
are solved for by extremizing a functional $\Gamma[G_R,G_K]$. In the exact formulation $\Gamma$ is given by
\begin{equation}
\Gamma \equiv g^{-1} G + M_{2\rm PI}[G].
\end{equation}
The functional $M_{2\rm PI}$ is the sum of all Feynman bubble diagrams that one may construct with electron lines and the four point vertex generated by the interaction,
excluding any diagrams which may be disconnected by cutting two electron lines (hence "two-particle irreducible").
Each diagram is then evaluated treating $G(t,t')$ as an unknown function to be determined by extremizing $\Gamma$.
As this involves a sum over all possible Feynman diagrams, it is not tractable.
Instead we approximate the sum $M_{2\rm PI}[G]$ by some subset of diagrams.

Here we take the $N\gg 1$ limit, which selects a string of bubble diagrams:
\begin{gather}
\Gamma[G] \approx g_R^{-1}(1,2)G(2,1) + \Pi(1,2)D(2,1),\\
D(1,2) \equiv [U^{-1} - \Pi]^{-1} (1,2),\\
\Pi(1,2) \equiv i G(1,2)G(1,2),
\end{gather}
where each of the functions $G$, $D$ and $\Pi$ has a $2\times 2$ Keldysh matrix structure. Note the effect of the $1/N$ expansion is to
select the Cooper interaction channel which is also the most singular channel near the critical point~\cite{Lemonik17}.

After having made the approximation to $\Gamma$, the Green's function is determined by setting $\delta \Gamma/ \delta G = 0$ which gives the Dyson equation
\begin{subequations}
\label{eq:EqOfMo}
\begin{equation}
{G_R}^{-1} = {g_R}^{-1}-\Sigma_R[G]; \,\, G_K = G_R\Sigma_KG_A,
\end{equation}
where $g^{-1}$ is the non-interacting Green's function and $\Sigma_{R,K}$ is the retarded, Keldysh self-energy.
In the present approximation $\Sigma_R$,
\begin{equation}
\Sigma_R(1,2) =  \frac{i}{N}\left[D_K(1,2)G_A(2,1) + D_R(1,2)G_K(2,1)\right],
\end{equation}
while $\Sigma_K$ is
\begin{eqnarray}
\Sigma_K(1,2) =  \frac{i}{N}\left[D_K(1,2)G_K(2,1) + D_R(1,2)G_A(2,1)+D_A(1,2)G_R(2,1)\right],
\end{eqnarray}
where $D$ is the Cooper fluctuation whose retarded and Keldysh parts are then
\begin{align}
D_R^{-1}&\equiv U^{-1}-\Pi_R,\label{eq:defDR}\\
D_K &\equiv D_R\circ \Pi_K \circ D_A. \label{eq:defDK}
\end{align}
Here $\Pi$ is the Cooper bubble. This is, in the present approximation, the expectation
\begin{align}
i\Pi^K(q,t,t')&= \langle \{\Delta_q(t),\Delta^\dagger_q(t')\}\rangle, \label{eq:defPiK}\\
i\Pi^R(q,t,t')&= \theta(t-t')\langle \left[\Delta_q(t),\Delta^\dagger_q(t')\right]\rangle.\label{eq:defPiR}
\end{align}
\end{subequations}
The function $D$ can be understood as the correlator of an auxiliary or Hubbard-Stratonovich field $\phi$ conjugate to $\Delta$,
used to decouple the fermionic quartic interaction in the Cooper channel~\cite{Lemonik17}.

The Keldysh component of the Dyson equation gives (where by $\circ$ we mean convolution),
\begin{align}
\biggl[i\partial_{t_1} -\epsilon\left(k + A(t_1)\right)+ i\partial_{t_2}& + \epsilon\left(k + A(t_2)\right)\biggr]G_K(k,t_1,t_2)\nonumber\\
&= \Sigma_K(k,t_1,t') \circ G_A(k,t',t_2) + \Sigma_R(k,t_1,t')\circ G_K(k,t',t_2)\nonumber\\
&{}-\biggl[G_R(k,t_1,t')\circ\Sigma_K(k,t',t_2) + G_K(k,t_1,t')\circ \Sigma_A(k,t',t_2)\biggr].\label{qkef1}
\end{align}

\subsection{Conservation laws}
We show how number conservation holds. For notational convenience we set $A= 0$. The conservation of energy and momentum can be shown to hold in a similar way.
It is helpful to substitute the expression for self-energies above, so that,
we obtain,
\begin{eqnarray}
&&\biggl[i\partial_{t}+ i\partial_{t''}\biggr]G_K(k,t,t'')\biggr \vert_{t''=t}=\nonumber\\
&&i\sum_q \int_{t'}\biggl\{  G_A(-k+q,t',t)G_A(k,t',t)D_R(q,t,t') +  G_K(-k+q,t',t)G_A(k,t',t)D_K(q,t,t') \biggr\} \nonumber\\
&&+i\sum_q \int_{t'}\biggl\{G_K(-k+q,t',t)G_K(k,t',t)D_R(q,t,t')+ G_A(-k+q,t',t)G_K(k,t',t)D_K(q,t,t')\biggr\} \nonumber\\
&&-i\sum_q \int_{t'}\biggl[\biggl\{G_R(k,t,t')G_R(-k+q,t,t')D_A(q,t',t) +  G_R(k,t,t')G_K(-k+q,t,t')D_K(q,t',t)\biggr\} \nonumber\\
&&+i\sum_q \int_{t'} \biggl\{G_K(k,t,t') G_R(-k+q,t,t')D_K(q,t',t)+ G_K(k,t,t')G_K(-k+q,t,t')D_A(q,t',t)\biggr\}\biggr].\label{qkef}
\end{eqnarray}
For number conservation we take the sum on $k$. The above terms can be further combined to give,
\begin{eqnarray}
&&\biggl[i\partial_{t}+ i\partial_{t''}\biggr]\sum_k G_K(k,t,t'')\biggr \vert_{t''=t}=
\sum_q\int_{t'}\biggl\{\Pi_K(q,t',t)D_R(q,t,t') + \Pi_A(q,t',t)D_K(q,t,t') \biggr\} \nonumber\\
&&- \sum_q\int_{t'}\biggl\{\Pi_K(q,t,t')D_A(q,t',t) +  \Pi_R(q,t,t')D_K(q,t',t)\biggr\}.
\end{eqnarray}
Now we use equations~\eqref{eq:defDR},~\eqref{eq:defDK}, so that
the second and fourth terms in the above equation get changed to,
\begin{eqnarray}
&&\biggl[i\partial_{t}+ i\partial_{t''}\biggr]{\rm Tr}\biggl[G_K(k,t,t'')\biggr]_{t''=t}=
{\rm Tr}\int_{t'}\biggl\{D_R(t,t')\Pi_K(t',t) + \biggl(D_R \Pi_K D_A\biggr)_{t,t'}\Pi_A(t',t) \biggr\} \nonumber\\
&&- \int_{t'}\biggl\{\Pi_K(t,t')D_A(t',t) +  \Pi_R(t,t')\biggl(D_R \Pi_K D_A\biggr)_{t',t}\biggr\}.
\end{eqnarray}
Writing $D_{A,R}\Pi_{A,R} = U^{-1}D_{A,R}-1$,
\begin{eqnarray}
\biggl[i\partial_{t}+ i\partial_{t''}\biggr]{\rm Tr}\biggl[G_K(t,t'')\biggr]_{t''=t}
=U^{-1}{\rm Tr}\biggl[ D_R \Pi_K D_A\biggr] - U^{-1}{\rm Tr}\biggl[D_R\Pi_K D_A\biggr] =0.
\end{eqnarray}
Thus the total particle number ${\rm Tr}\left[G_K\right]$ is a constant of motion.

\section{Reduction to a quasi-classical kinetic equation}

There are three main frequencies of interest in the problem: (i)  the Fermi energy $E_F$, (ii) the temperature $T$, (iii)
the rate of collisions between the fermions and Cooper fluctuations $1/\tau_{\rm sc}$.
In the systems of physical interest, $E_F \gg T$. In addition, as long as the system is not too close to the critical point,
$T \gg \tau^{-1}_{\rm sc}$. If $T \gg \tau_{\rm sc}^{-1}$ then we do not need to consider repeated coherent scattering off 
of the superconducting fluctuations. 
As the the scattering rate will increase with size of the superconducting fluctuations 
we must give an upper limit to the validity of neglecting repeated scattering. 

The Green's functions always appear in combinations as in $\Pi$. Let us first evaluate $\Pi$ in the non-interacting, $E=0$ limit,
\begin{align}
\Pi_R(q;t_1,t_2) &= i\sum_k G_K(k +q/2,t_1,t_2)G_R(-k +q/2,t_1,t_2)\nonumber\\
			&= -i\sum_k n(k+ q/2)e^{-i(t_1 - t_2)(\epsilon(k+\frac{q}{2}) + \epsilon(k -\frac{q}{2}))}.
\end{align}

As the function $n(k + q/2)$ is smooth on scales smaller than $T/v$, $v$ being the average Fermi velocity, this sum should decay exponentially as $\exp\left[-T (t_1 - t_2)\right]$. Thus we need only
concern ourselves with times $T(t_1 -t_2)$ not much larger than $1$.  Therefore if the self-energy $\Sigma_R$ is $\ll T$, it will only perturbatively
correct this exponential decay. The precise form of $\Sigma_R$ is discussed in~\cite{Lemonik17} for the hard quench, and we repeat a similar calculation here. 
To give a precise criterion for the size of the fluctuations, we note that within the 2PI approximation, for large fluctuations we must solve self-consistently the equations,
\begin{align}
\Sigma_R(k,\omega) &= \int \frac{d^2q}{(2\pi)^2} \frac{i D_K(q)}{-\omega - \varepsilon_{-k+q} -\Sigma_A(-k, -\omega)},\\
\Sigma_A(-k,-\omega) &= \int \frac{d^2q}{(2\pi)^2} \frac{i D_K(q)}{\omega - \varepsilon_{k+q} -\Sigma_R(k, \omega)}.
\end{align}
Defining $z = -\omega + \epsilon  + \Sigma_R(\epsilon,\omega)$ and $z' = \omega + \epsilon +\Sigma_A(\epsilon,-\omega)$, this is equivalent to the self-consistent pair of equations
(in $d=2$),
\begin{align}
z &= -\omega + \epsilon + \int \frac{d( q^2)}{4\pi^2} \frac{i D_K(q)}{\sqrt{v^2q^2 - (z')^2}} \\
z' &= \omega + \epsilon + \int \frac{d( q^2)}{4\pi^2} \frac{i D_K(q)}{\sqrt{v^2q^2 - (z)^2}} 
\end{align}
We now solve these equations recursively starting from the free solutions, $z_0 =  -\omega + \epsilon + i\delta$, $z_0' = \omega + \epsilon -i\delta $
\begin{align}
z_1 &= -\omega + \epsilon + \int \frac{d( q^2)}{4\pi^2} \frac{i D_K(q)}{\sqrt{v^2q^2 - (z'_0)^2}} \\
z_2 &= -\omega + \epsilon + \int \frac{d( q^2)}{4\pi^2} \frac{i D_K(q)}{\sqrt{v^2q^2 - (z'_1)^2}}
&+\text{ etc...}
\end{align}
Considering the most divergent case when $\epsilon = \omega = 0$, let us calculate the difference between the first and second terms of the sequence,
\begin{align}
(z_1 - z_2)/z_1 &=  \frac{1}{z_1}\int \frac{d( q^2)}{4\pi^2} i D_K(q)\left[ \frac{1}{\sqrt{v^2q^2 }} - \frac{1}{\sqrt{v^2q^2 - (z'_1)^2}} \right]\\
&\approx\frac{1}{z_1}\int \frac{d( q^2)}{4\pi^2} i D_K(q = 0 )\left[ \frac{1}{\sqrt{v^2q^2 }} - \frac{1}{\sqrt{v^2q^2 - (z'_1)^2}} \right]\\
&\propto \frac{1}{z_1} \cdot\frac{1}{4\pi^2}\left(iD_K(q =0) \right)\cdot\frac{z'_1}{v^2}\\
&\propto \frac{F(q=0)T}{\nu v^2} \approx F(q =0)\cdot\frac{T}{E_F}
\end{align}
In the second line we use the fact that the term in the brackets is more sharply peaked around $q=0$ than $i D_K(q)$  
so only the constant part of $iD_K$ needs to be retained.  If the final term $F(0)T/E_F$ is much less than one, 
than we may stop with the first approximation $z_1$. That is the self energy does not need to be self-consistently 
included within with the diagrams we consider.

Therefore it is sufficient  in all regimes to replace $G_R\rightarrow g_R$ on the right-hand side of the kinetic equation. Having made this approximation we may now determine the behavior of
$G_K(t_1,t_2)$, for $(t_1 -t_2)T$ small. Starting with $G_K = G_R\Sigma_K G_A$ and left convolving with $G_R^{-1}$ we obtain,
\begin{align}
(i\partial_{t_1} - \epsilon_k(t_1))G_K(t_1,t_2) &= \int dt' \bigg[\Sigma_R(t_1,t')G_K(t',t_2) +\Sigma_K(t_1,t')G_A(t',t_2)\bigg].
\end{align}

Again the RHS can be neglected if $t_1 - t_2$ is small compared to the $\Sigma^{-1}_R$. In this case, the behavior of $G_K$ is a simple exponential
$\propto e^{-i \epsilon_k(t_1 - t_2)}$. Assuming temporarily that we know the distribution function $n_k(t)$ at all times,
we may choose the boundary condition $iG_K(t,t)  = n_{k+ A(t)}(t)$,
\begin{equation}
G_K(k,t_1,t_2) = g_R(k,t_1,t_2)n_{k+A(t_2)}(t_2) - n_{k+A(t_1)}(t_1)g_A(t_1,t_2).
\end{equation}
With these approximations the polarization $\Pi$ is given as a functional of $n_k(t)$,
\begin{align}
\Pi_R(q,t_1,t_2) &= \frac{i}{2}\sum_k g_R(k+q/2;t_1,t_2)g_R(-k+q/2;t_1,t_2)\biggl[n_{k+q/2 +A(t_2)}(t_2)+n_{-k+q/2 + A(t_2)}(t_2)\biggr]\nonumber\\
&=\sum_k\Pi_R'(k,q,t_1,t_2),\nonumber \\
\Pi_K(q,t_1,t_2) &= \frac{i}{2}\sum_k g_R(k+q/2;t_1,t_2)g_R(-k+q/2;t_1,t_2)\nonumber\\
					&\quad{}\times\biggl[n_{-k+q/2 + A(t_2)}(t_2)n_{k+q/2 + A(t_2)}(t_2) +1\biggr]\nonumber \\
					&{}+ \frac{i}{2}\sum_k g_A(k+q/2;t_1,t_2)g_A(-k+q/2;t_1,t_2)\nonumber\\
					&\quad{}\times\biggl[n_{-k+q/2 + A(t_1)}(t_1)n_{k+q/2 + A(t_1)}(t_1 ) +1\biggr]\nonumber \\
					&= \sum_k\biggl[\Pi_K'(k,q,t_1,t_2)\theta(t_1 - t_2) + \Pi_K'(k,q,t_1,t_2)\theta(t_2 - t_1) \biggr],
\end{align}
where the above equation also provide the definition of $\Pi'$.
Note that when the equilibrium distribution function
$n_k(t) = n_k^{\rm eq}=\text{tanh}(\frac{\epsilon_k}{2T})$ is inserted in this expression, the fluctuation dissipation theorem holds.

Evaluating the kinetic equation~\eqref{qkef1} at equal times $t_1=t_2=t$, inserting the definition
$ n_k(t) = iG_K(k - A(t), t,t) $ on the LHS, shifting $q\rightarrow q-2A(t)$ for gauge invariant results,
and employing the above approximations on the RHS we obtain the kinetic equation,
\begin{align}
\partial_t n_k(t) - \vec{E}\cdot\vec{\nabla}_k n_k(t)  &= \sum_q\,\int_0^t\!\!dt' \text{Im}\biggl[
	iD_K(q-2A(t),t,t')\Pi_A'(k -A(t),q - 2 A(t),t',t)
	\nonumber\\
	&+
iD_R(q-2A(t),t,t') \Pi_K'(k - A(t),q - 2 A(t),t',t)\biggr].
\end{align}
This must be supplemented by the equations of motion for $D$ which are the same as~\eqref{eq:defDR},~\eqref{eq:defDK} but where
 $\Pi$ is calculated in the present approximation. Conservation of charge is then ensured as long as
\begin{equation}
\sum_k \Pi'(k,q; t_1,t_2) = \Pi(q;t_1,t_2),
\end{equation}
as may be seen by following directly the earlier derivation.

\subsection{Linear response to an electric field}
The previous derivation is correct for arbitrary electric fields strengths, and includes non-linear effects. We now expand in powers of $E(t)$. As $\delta n$ and $D$ vary at rates much slower than $T$, it is  sufficient to calculate the $\Pi$
to lowest order in frequency giving~\cite{Lemonik17},
\begin{align}
\left(\partial_t  + \lambda_q\right)D_R &= T\nu^{-1}\delta(t-t');\,\,\,\lambda_q = r+ q^2/2M .\\
&\Rightarrow D_R(t,t') = \theta(t-t')T\nu^{-1}e^{-\lambda_q(t-t')},\\
iD_K(t,t') &= T\nu^{-1}\frac{T}{\lambda_q}\left(e^{-\lambda_q|t-t'|} - e^{-\lambda_q(t+t')}\right)\\
&= D_R(t,t')F(t') + F(t)D_A(t,t')\label{DKdef},\\
F(t)&\equiv\frac{T}{\lambda_q}\left(1 - e^{-2\lambda_qt}\right).
\end{align}
Note the function $F(t)$ changes with rate $\lambda_q$ which for small $q$ is $\ll T$.

We now expand the Dyson equation to linear order in $E(t)$ in order to evaluate the linear response. The leading change in the electron Green's functions are,
\begin{align}
g_R(t_1,t_2) &= -i\theta(t_1-t_2)e^{-i\epsilon_k(t_1 - t_2)}\left(1 -i \int_{t_2}^{t_1} \!\!ds v_k \cdot A(s)\right),\\
n_k &= n^{\rm eq}_k + \delta n_k(t),
\end{align}
where $\delta n$ is assumed to be of order $E$.
In this case the kinetic equation may be rewritten as
\begin{align}
&\partial_t n_k(t) - \vec{E}\cdot \vec{\nabla}_k n^{\rm eq}_k =
	\sum_q\int\!\!dt' \text{Im}\bigg[
	 iD_K(q-2A(t),t,t')\delta\Pi_A'(k-A(t),q-2A(t),t',t)\nonumber\\
	& + i D_R(q-2A(t),t,t')\delta\Pi_K'(k-A(t),q-2A(t),t',t)
	 \nonumber \\
	& +  \delta iD_K(q-2A(t),t,t')\Pi_A'(k-A(t),q-2A(t),t',t) +
	 i \delta D_R(q-2A(t),t,t')\Pi_K'(k-A(t),q-2A(t),t,t')
	 \biggl],
\end{align}
where the $\delta$ indicates the first order variation with respect to the electric field. As expected the conservation laws continue to hold after this approximation.
Since we are interested in the response of the current, we multiply with $v_k$ and integrate over $k$.
The fluctuations are modified by the electric field through their dependence on $\Pi_R$:
\begin{align}
\delta D_R &= D_R\circ \delta \Pi_R \circ D_R, \\
D_K& = \delta D_R \circ \Pi_K \circ D_A  + D_R \circ \delta\Pi_K \circ D_A + D_R \circ \Pi_K \circ \delta D_A.
\end{align}

\subsection{Single mode projection}
The kinetic equation is an integral-differential equation in which the unknowns $\delta n_k(t)$ appear linearly.
Conceptually, therefore it may be solved by standard methods.
However, even though it is linear, it is non-local in time and not time translation invariant.
This makes direct analytical and numerical solution difficult.

We therefore make an additional simplification that instead of considering the full space of solutions, we instead project entirely onto the current mode. This means that we would like to fix,
\begin{align}
\delta n_k(t)  \overset{?}{=} \frac{\bar{m}}{\rho} J^i(t) \nabla_k^i n^{\rm eq}_k ,\\
\qquad m^{-1}_{ij} = \frac{\partial^2 \epsilon(k)}{\partial k^i \partial k^j}=\partial_{k_j}v_k^i;\,\,
\bar{m}^{-1}\delta_{ij} = -\frac{1}{\rho}\sum_{k} m^{-1}_{ij}n^{\rm eq}_k
,
\label{eq:wrongJn}
\end{align}
where $\rho$ is the density of fermions,
$\bar{m}$ is the effective mass and $J^i(t)$ is some function to be determined. However because of the presence of conservation of momentum,
this approximation fails qualitatively. As $\partial_t P = \rho E $ by conservation of momentum, any initial perturbation
\begin{equation}
\delta n =  V^i k^i \partial n^{\rm eq}/\partial \epsilon,
\end{equation}
with $V^i$ arbitrary, will never decay.
Therefore instead of Eq.~\eqref{eq:wrongJn} we decompose the occupation number as
\begin{align}
\delta n_k(t) &= \frac{\bar{m}}{\rho}
	\left[J_r^i(t)( v_k^i - \gamma k^i/\bar{m})  +  \gamma k^i P^i(t)/\bar{m}^2\right]\partial_\epsilon n^{\rm eq}_k,\\
\gamma &\equiv \frac{\bar{m} \sum_k v_k\cdot k \partial_\epsilon n}{\sum_k k\cdot k\partial_\epsilon n},\\
J_{\rm tot}^i &\equiv \sum_k v_k^i \delta n_k = (1-\gamma)J_r^i + \gamma P^i/\bar{m}.
\end{align}
The parameter $\gamma$ gives the amount of current that is carried by the momentum mode, which does not relax. In the limit $\gamma \rightarrow 0$ there
is no overlap between the modes, the momentum mode carries no current, and is thus neglected. On the other hand,
when $\gamma \rightarrow 1$, as in a Galilean invariant system, the current and momentum are proportional and there is no relaxing current.

Multiplying the kinetic equation by $v_k - \gamma k/ \bar{m}$ and $k$, and summing over $k$ we obtain,
\begin{align}
\partial_t J^i_r(t) - \frac{\rho}{\bar{m}}E^i &=
	\sum_{k,q}
	\frac{v_k^i -\gamma k^i/m}
		 {1-\gamma}
	\int\!\!dt' \text{Im}\bigg[
	 iD_K(q - 2A(t),t,t')\delta\Pi^A(k - A(t),q- 2A(t) ,t',t)\nonumber\\
	 &+ i D_R(q- 2A(t) ,t,t')\delta\Pi^i_K(k-A(t),q- 2A(t), t',t)
	 \nonumber \\
	& +  \delta i D_K(q-2A(t),t,t')\Pi^i_A(k-A(t),q-2A(t),t',t)\nonumber\\
	 &+	 i \delta D_A(q- 2A(t),t,t')\Pi^i_K(k-A(t),q-2A(t),t,t')
	 \bigg],\\
\partial_t P^i(t) - \rho E^i(t) &= 0.
\end{align}
As the momentum mode has trivial dependence, we  will simply drop it and set $\gamma = 0$. In this limit,
\begin{eqnarray}
\delta n_k(t) &= \frac{\bar{m}}{\rho}J_r^i(t)\nabla^i_kn_k^{\rm eq}.\label{defn2}
\end{eqnarray}
These are now a closed set of equations in terms of the single unknown function $J_r(t)$ which we will denote simply
as $J(t)$. The remaining step is to evaluate the various terms.
We note that $D_K(t,t')$ is generally larger than $D_R$ by the factor $T/\lambda_q$. Thus we will only keep the terms that are highest order in $D_K$.
\section{Evaluation of kinetic coefficients}
We begin by evaluating $\delta\Pi^i_R(q,t,t')$. There are two contributions. The first is from varying $n_k$, and the second is from varying $g_R$:
\begin{align}
\text{Im}\int \!\!dt'\,\sum_k  v_k^i i D_K(q-2A(t), t,t') \delta\Pi^i_A(k -A(t),q -2A(t),t',t)&= K_J^{i}(t) + K_E^{i}(t).
\end{align}
The first term, obtained from varying $n_k$ is,
\begin{align}
K_J^{i}(t) \equiv \frac{1}{2}\text{Im}\int \!\!\sum_{k,q}	iD_K(q,t,t')(i)
	\left(
		\delta n_{k+q/2}(t') + \delta n_{-k +q/2}(t')
	\right)
	\left(
		v^i_{k+q/2} + v^i_{-k + q/2}
	\right)
	e^{i\left(
		\epsilon_{k+q/2} + \epsilon_{-k+q/2}
	\right)(t-t')}.
\end{align}
Substituting Eq.~\eqref{DKdef} for $D_K$, we see that this can be rewritten as
\begin{align}
K_J^{i}(t) &=  \int_0^t dt' \sum_q Q^{ij}_J(t,t') F_q(t')J^{j}(t'),
\nonumber\\
Q^{ij}_J(t,t') &\equiv
	\frac{\bar{m} T }{2 \rho \nu}\text{Im}\int dt' \sum_{k}
	\left(
		\nabla^j_k n^{\rm eq}_{k+q/2}
		+
		\nabla^j_k n^{\rm eq}_{-k+q/2}
	\right)
	(i)
	\left(
		v^i_{k+q/2}
		+
		v^i_{-k + q/2}
	\right)
	e^{i\left(
		\epsilon_{k+q/2} + \epsilon_{-k+q/2} + i\lambda_q
		\right)(t-t')
		}.
		&
\end{align}
Similarly, the change in $g_R$ due to the electric field gives,
\begin{align}
K_E^{i}(t) &=  \int_0^t dt' \sum_q Q^{ij}_E(t,t')F_q(t') E^j(t'),
\nonumber\\
Q^{ij}_E(t,t')E^j(t') &\equiv-\text{Im}
	\frac{T}{2\nu} \sum_{k}
	\left(
		n^{\rm eq}_{k+q/2}
		+
	    n^{\rm eq}_{-k+q/2}
	\right)
	\left(
		v^i_{k+q/2}
		+
		v^i_{-k + q/2}
	\right)\nonumber\\
	&\quad{}\times\int_{t'}^{t}\! ds\,v_k^j\left( A^j(t) - A^j(s)\right)
	e^{i\left(
		\epsilon_{k+q/2} + \epsilon_{-k+q/2} + i\lambda_q
		\right)(t-t')
		}.
\end{align}

Following the previous discussion the terms $Q^{ij}_{E,J}$ decay exponentially with rate $T$, whereas $J$ and $F$ change at a slower rate.
Therefore $Q_{E,J}$ appear like delta functions when integrated against $F(t)J(t)$, and we may write
\begin{align}
K_J^i(t) &\approx \sum_q F_q(t)J^j(t)\int_0^t dt'Q^{ij}_J(t,t'),\nonumber\\
\int_0^t dt'Q^{ij}_J(t,t') &\approx
	-\frac{\bar{m} T }{2\rho\nu}\text{Im} \sum_{k}
	\frac{
	\left(
		\nabla^j_k n^{\rm eq}_{k+q/2}
		+
		\nabla^j_k n^{\rm eq}_{-k+q/2}
	\right)
	\left(
		v^i_{k+q/2}
		+
		v^i_{-k + q/2}
	\right)
	 }{
		\epsilon_{k+q/2} + \epsilon_{-k+q/2} + i\lambda_q
		}
	\nonumber\\
	&\approx_{v q \ll T}-
	\frac{\bar{m} T }{2\rho\nu}\text{Im} \sum_{k}
	\frac{
		q^r q^l m^{-1}_{il}\nabla^r_k\nabla^j_k n^{\rm eq}_{k}
	}{
		2\epsilon_{k} + i\lambda_q
		}
	\nonumber\\
	&\approx_{\lambda_q \ll T}
		\frac{\pi \bar{m} T}{2\rho\nu} \sum_{k}
	\left(
		q^r q^l m^{-1}_{il}\nabla^r_k\nabla^j_k n^{\rm eq}_{k}
	\right)
	\delta(\epsilon_{k} + \epsilon_{-k})
\nonumber\\
	&= \frac{\pi \bar{m} T}{4\rho\nu} \sum_{k}
	\left(
		q^r q^l m^{-1}_{il}\frac{\partial^2\epsilon_k}{\partial_k^r\partial_k^j}\partial_{\epsilon_k}n^{\rm eq}_{k}
	\right)
	\delta(\epsilon_{k})
\nonumber\\
	&\approx
	\frac{\bar{m}\pi q^r q^l }{4  \rho}\langle m_{jr}^{-1} m_{il}^{-1} \rangle_{\rm FS}.
\end{align}
In the second line we assume $t^{-1} \gg T$ so we are not too close to the quench. In the third line we take $q << T/v$, and we also assume $\lambda_q/T \ll 1$.
To estimate the magnitude of this term, we neglect factors of order one, write $\rho \sim \nu k^2_F/m$, and obtain that this term is $\sim (q/k_F)^2/\nu $.

We proceed similarly for the electric field. Assume that the frequency of the electric field $\omega \ll T$ - in this case we may approximate the integral
\begin{equation}
\int_{t'}^{t} ds (A^j(t) - A^j(s)) \approx -E^j(t)(t-t')^2/2.
\end{equation}
Approximating as before for $Q^{ij}_J$, we have
\begin{align}
K_E^i(t) &\approx \sum_q F_q(t)\frac{\rho E^j(t)}{\bar{m}}\int_0^t
dt'Q^{ij}_E(t,t'),\nonumber\\
\int_0^t dt'Q^{ij}_E(t,t') &\approx
	\frac{\bar{m} T}{2\rho \nu}\text{Im}\sum_{k}i
	\frac{
	\left(
		n^{\rm eq}_{k+q/2}
		+
		n^{\rm eq}_{-k+q/2}
	\right)
	\left(
		v^i_{k+q/2}
		+
		v^i_{-k + q/2}
	\right)\left(
		v^j_{k+q/2}
		+
		v^j_{-k + q/2}
	\right)
	 }{
		(\epsilon_{k+q/2} + \epsilon_{-k+q/2} + i\lambda_q)^{3}
		}
	\nonumber\\	
	&\approx
	\frac{\bar{m} T}{2\rho \nu}\text{Im}\sum_{k}i
	\frac{
		q^l q^r m^{-1}_{il} m^{-1}_{jr}n^{\rm eq}_{k}
	}{
		(2\epsilon_{k} + i\lambda_q)^3
		}
	\nonumber\\
 	&\approx
    \frac{\pi \bar{m} q^r q^l}{2 \rho}\langle m_{jr}^{-1} m_{il}^{-1} \rangle_{FS} \int dx x^{-1} \frac{d^2}{d x^2 } \tanh(x/2)\nonumber\\
    &\approx
    \frac{14\pi \bar{m}  q^r q^l}{ \rho}\zeta'(-2)\langle m_{jr}^{-1} m_{il}^{-1} \rangle_{\rm FS}.
\end{align}
The resulting term has the form $\propto (q/k_F)^2 F_q(t) \rho E(t)/\bar{m}$, and is thus a parametrically small correction to the drift term.
Therefore we neglect it for the remainder.

We now turn to the second term $\sim\delta D_K \Pi'_A$. We begin by considering the change in $\delta D_K$.  This in turn depends on $\delta D_R$,
which is given by
\begin{equation}
\left[\partial_t + \lambda_q\right]\delta D_R = -\delta \lambda_q D_R.
\end{equation}

There are two contributions to $\delta\lambda_q$. One is from varying $\delta n$ via
\begin{equation}
\lambda_q = U^{-1} -{\rm Re} \biggl[\Pi_R(q)\biggr]=U^{-1} +  \sum_k \frac{n_{k+q/2}+n_{-k+q/2}}{\epsilon_{k+q/2}+\epsilon_{-k+q/2}}.
\end{equation}
Thus,
\begin{align}
\delta \lambda_q &=
\sum_k\frac{\delta n_{k+q/2}+\delta n_{-k+q/2}}{\epsilon_{k+q/2}+\epsilon_{-k+q/2}}
\nonumber\\
&= \sum_k\frac{\delta n_{k+q/2}-\delta n_{k-q/2}}{\epsilon_{k+q/2}+\epsilon_{-k+q/2}}\nonumber\\
&\simeq q^i\sum_k\frac{\nabla^i_k\delta n_{k}}{2\epsilon_{k}}\qquad (q\rightarrow 0 ).
\end{align}
as $\delta n_k =  -\delta n_{-k}$. Using the relation between $\delta n_k$ and the current,
\begin{align}
\delta \lambda_q &= \frac{q^i \bar{m}}{\rho} J^j\sum_k \frac{\nabla^i_k\nabla^j_k n^{\rm eq}_k}{2\epsilon_k}\nonumber\\
	&= 4\frac{q^i \bar{m}}{\rho} J^j \frac{\partial^2}{\partial q ^i \partial q^j} \lambda_q = 4\frac{\bar{m}}{\rho M}q^i J^i,
\end{align}
where we have used the definition $\lambda_q = r+ q^2/(2M)$, with $M\sim T/v^2$.

The second reason for the change in $\lambda_q$ is due to the direct coupling to the electric field, where defining $\delta A(s) = A(s) -A(t)$,
\begin{eqnarray}
&&\lambda_q(s) = \frac{(q+2\delta A(s))^2}{2M} \Rightarrow \delta \lambda_q = \frac{2 q \delta A(s)}{M} = \frac{2q}{M}\left[A(s)-A(t)\right]=
-\frac{2q}{M}\int_s^t dt' \partial_{t'} A(t'),\nonumber\\
&&\Rightarrow \delta \lambda_q= \frac{2q}{M}\int_s^t dt' E(t').
\end{eqnarray}
Thus the total change in $\lambda_q$ is
\begin{align}
\delta\lambda_q(s) = \frac{ 4 \bar{m} q^iJ^i(s)}{ M \rho} +\frac{2q^i}{M}\int_s^t dt' E^i(t')=
\frac{4 \bar{m} q^i}{ \rho M}\left[J^i(s)+ \frac{1}{2}\int_s^t dt' \frac{\rho E^i(t')}{\bar{m}}\right].
\end{align}
Now we use,
\begin{equation}
\delta D_R(t,t' ) = e^{-\lambda_q(t-t')}
	\left[
		-\int_{t'}^t ds \delta \lambda_q(s)
	\right].
\end{equation}

Note that $\delta D_K = (\delta D_R) \Pi_K D_A+ D_R(\delta \Pi_K) D_A+D_R \Pi_K (\delta D_A)$.
Since the variation $\delta \Pi_K$ produces a term that is smaller by a factor of $F_q$, we neglect it. Using the zeroth order calculation that
$i\Pi_K = \nu \delta(t -t')$, we thus have,
\begin{equation}
i\delta D_K(t , t') = \nu \int ds \delta D_R(t,s) D_A(s, t') + \nu \int ds  D_R(t,s) \delta D_A(s, t').
\end{equation}
Plugging in the result for $\delta D_R(t,t') = \delta D_A(t',t)^*$, we obtain for the case $t > t'$
\begin{align}
i\delta D_K(t , t') &= -\nu  \int_0^{t'} ds D_R(t,s) D_A(s,t')\bigg[ \int^{t}_s du \delta \lambda_q(u) + \int^{t'}_s du \delta \lambda_q(u) \bigg] \\
	&= -2\nu\int^{t'}_0 du \delta \lambda_q(u) \int^u_0 ds D_R(t,s) D_A(s,t')
	 -\nu\int^t_{t'}du\delta \lambda_q(u) \int_0^{t'} ds D_R(t,s) D_A(s,t').
\end{align}
Using the definitions of $D_K$ and $D_R$ this is,
\begin{align}
i\delta D_K (t> t') &= -\frac{2T}{2\nu}e^{-\lambda_q(t-t')} \int_0^{t'}\!du\, e^{-2\lambda_q(t' - u)}  \delta \lambda_q(u) F_q(u)
-\frac{T}{2\nu} e^{-\lambda_q(t-t')} \int_{t'}^t du\delta\lambda_q(u) F_q(t')
\nonumber\\
&=
	-\frac{2T}{2\nu} e^{-\lambda_q(t-t')}
		\int_0^{t}\!du\, e^{-2\lambda_q(t' - u)}
		\delta \lambda_q(u) F_q(u)
    -\frac{T}{2\nu} e^{-\lambda_q(t-t')}
    	\int_{t'}^t du
    	\delta\lambda_q(u) \left(F_q(t') - 2F_q(u)\right)
\nonumber\\
&=
	-\frac{2T}{2\nu} e^{\lambda_q(t-t')}
		\int_0^{t}\!du\, e^{-2\lambda_q(t - u)}
		\delta \lambda_q(u) F_q(u)
    -\frac{T}{2\nu} e^{-\lambda_q(t-t')}
    	\int_{t'}^t du
    	\delta\lambda_q(u) \left(F_q(t') - 2F_q(u)\right)
\nonumber\\
&\approx -\frac{2T}{2\nu} e^{\lambda_q(t-t')}
		\int_0^{t}\!du\, e^{-2\lambda_q(t - u)}
		\delta \lambda_q(u) F_q(u)
    +\frac{T}{2\nu} (t-t')
    	\delta\lambda_q(t)F_q(t),
\end{align}
where in the last line we used the fact that $t-t'$ is of order $T^{-1}$. We discuss this last term below and
show it to be parametrically smaller than the local terms calculated earlier.

We insert the second part into the collision integral giving,
\begin{align}
\int_0^t dt'\text{Im} \Pi^i_A(t',t)
	&\left(
		e^{-\lambda_q(t-t')}
		\frac{T}{\nu} (t-t') \delta \lambda_q(t) F_q(t)
	\right)
\nonumber\\
		 &=
	\delta\lambda_q(t) F_q(t)
	\text{Im}\sum_k
	i \frac{
		n^{\rm eq}_{k+q/2} + n^{\rm eq}_{-k+q/2}
	}{\
		(\epsilon_{k+q/2} +\epsilon_{-k +q/2} - i\lambda_q)^2
	}
\nonumber\\
	&\approx_{q\rightarrow 0}
	\delta\lambda_q(t) F_q(t)
	\text{Im}	
	\sum_k i
	\frac{
		2n(\epsilon_k)
	}{
		(2\epsilon_k - i \lambda_q)^2
	}
\nonumber\\
	&\approx
	\delta\lambda_q(t) F_q(t)
	\text{Im}	
	i\int d\epsilon \, \nu
	\frac{
		2n(\epsilon)
	}{
		(2\epsilon - i \lambda_q)^2
	}
\nonumber\\
	&\approx_{\lambda\rightarrow 0}
	\delta\lambda_q(t) F_q(t)
	\text{Re}	
	\int d\epsilon \, \nu
	\mathcal{P}\frac{
		\partial_\epsilon n(\epsilon)
	}{
		\epsilon
	} = 0.
\end{align}
Thus this term is parametrically smaller than the local term already calculated.

Now we consider the full term
\begin{align}
\int_0^{t} dt' \text{Im}&\left[v_k^i i \delta D_K(t,t') \Pi'_A(t',t)
\right]
 \nonumber\\
 &=  \frac{1}{2}\text{Im}\int_0^t dt'
 \sum_{k,q}(i)
 \left(v^i_{k+q/2} + v^i_{-k+q/2}\right)
 \left(n^{\rm eq}_{k+q/2}+ n^{\rm eq}_{-k+q/2}\right)
 e^{i(\epsilon_{k+q/2} + \epsilon_{-k+q/2})(t-t')}
 i\delta D_K(t',t)
 \nonumber\\
&\approx
 \text{Im} \frac{T}{2\nu}
 \sum_{k,q} (i)\int_0^t dt'
 \left(v^i_{k+q/2} + v^i_{-k + q/2}\right)
 \left(n^{\rm eq}_{k+q/2}
+ n^{\rm eq}_{-k+q/2}\right)
e^{i(\epsilon_{k+q/2} + \epsilon_{-k+q/2} -i \lambda_q)(t-t')}
 \nonumber\\
&\quad{}\times\int_0^{t} du\,
\left(-
 e^{-2(t-u)\lambda_q}
 \delta\lambda_q(u) F_q(u)
 \right)
\nonumber \\
&\approx
 \text{Im}\frac{T}{2\nu}\sum_{k,q}
 q^j m^{-1}_{ij}
 \frac{n^{\rm eq}_{k+q/2} + n^{\rm eq}_{-k+q/2}}
      {\epsilon_{k+q/2} + \epsilon_{-k+q/2} - i\lambda_q}
 \int_0^t du
 \,e^{-2(t-u)\lambda_q}
 \delta \lambda_q(u) F_q(u)
\nonumber\\
& \approx\sum_q
	\frac{ q^l T}{2\nu}
	\sum_k m^{-1}_{il}
	\frac{n^{\rm eq}(\epsilon_k) \lambda_q}
	     {4\epsilon_k^2 + \lambda_q^2}
    \int_0^t du
    e^{-2(t-u)\lambda_q}
    \delta \lambda_q(u) F_q(u)
\nonumber\\
& = \sum_q
	\frac{T q^l }{2\nu}
	\left(
		\sum_k m^{-1}_{il}
		\frac{n^{\rm eq}(\epsilon_k) \lambda_q}
		     {4\epsilon^2_k + \lambda_q^2}
	\right)
	\int_0^t du
	e^{-2(t-u)\lambda_q}
	F_q(u)
\nonumber\\
&\quad{}\times
	\frac{4 q^j\bar{m}}{\rho M}
	\left[
		J^j(u) +
		\frac{1}{2}\int_u^t dt'
		\frac{\rho E^j(t')}
			 {\bar{m}}
	\right]
\nonumber\\
& = \sum_q\alpha q^i q^j \lambda_q
	\int_0^t du
	e^{-2(t-u)\lambda_q} F_q(u)
	\left[
		J^j(u)+
		\frac{1}{2}\int_u^t dt'
		\frac{\rho E^j(t')}
			 {\bar{m}}
	\right].
\end{align}
The coefficient $\alpha$ is given by,
\begin{align}
\alpha &\equiv \frac{2 T \bar{m} }{M \nu \rho} \sum_k m^{-1}_{ii}\frac{n^{\rm eq}(\epsilon_k)}{4\epsilon_k^2 + \lambda_q^2}\nonumber\\
&= \frac{2 T \bar{m} }{M \nu \rho} \int\!\! d\epsilon\, \kappa(\epsilon)\frac{n^{\rm eq} }{4\epsilon^2 + \lambda^2}\nonumber \\
\kappa(\epsilon) &\equiv  \sum_k m^{-1}_{ii}(k)\delta(\epsilon - \epsilon_k).
\end{align}
The function $\kappa(\epsilon)$ may be Taylor expanded to give $\kappa(\epsilon) \approx  \frac{\nu}{\bar{m}}(1 + b \epsilon/E_F + \cdots)$, where the parameter $b$ is a material parameter of order $1$. The sign of $b$ is not fixed in general.  However for a 'simple' band structure, the parameter $b$ is negative. Inserting this into the expression for $\alpha$ gives

\begin{align}
\alpha &\approx \frac{ 2 T }{M \rho} \int d\epsilon \left(1+b\frac{\epsilon}{E_F}\right) \frac{n^{\rm eq}(\epsilon) }{4\epsilon^2 + \lambda^2}\nonumber\\
		&= \frac {2 b T}{M\rho E_F} \int d\epsilon \frac{\epsilon n^{\rm eq}(\epsilon)}{4\epsilon^2}\nonumber\\
&\approx \frac{b T\log(E_F/T)}{M\rho E_F} \nonumber \\
&\approx \frac{b v^2_F T\log(E_F/T)}{T\nu E_F^2} \nonumber\\
&\approx \frac{1}{k_F^2 \nu} b\log(E_F/T).
\end{align}
We have that $\alpha q^2 \sim (q/k_F)^2/\nu$, which is comparable to the local terms. Moreover the sign of $\alpha$ for generic
band structures is
such as to oppose the  local in time term coming from $iD_K\delta\Pi_A$.

Combining the results, and setting $\bar{m}/\rho = 1$ for convenience
\begin{align}
\partial_t J^i = E^i(t) - A(t)J^i(t) +\alpha \sum_q q^2 \lambda_q\int_0^t ds\biggl[J^i(s) +\frac{1}{2}\int_s^t dt' E^i(t')\biggr]
e^{-2\lambda_q(t-s)}F_q(s).\label{kinfin}
\end{align}
We write this as
\begin{align}
\partial_tJ = E(t) - A(t)J^i(t) +\alpha \int_0^t ds B(t,s)\biggl[J(s)+\frac{1}{2}\int_s^t dt' E(t')\biggr].\label{J1a}
\end{align}
Writing $B(t,s) = \frac{d}{ds}C(t,s), C(t,0)=0$, or $C(t,s) = \int_0^s ds' B(t,s')$. Then,
\begin{align}
\int_0^t ds B(t,s) \int_s^t dt' E(t') =\int_0^t ds  \frac{d}{ds}C(t,s) \int_s^t dt' E(t')=\int_0^tds C(t,s)E(s).
\end{align}
Thus the kinetic equation is,
\begin{align}
\partial_tJ = E(t) - A(t)J(t) +\alpha \int_0^t ds \biggl[B(t,s)J(s) + \frac{1}{2}C(t,s)E(s)\biggr].\label{J1}
\end{align}

Above,
\begin{align}
A(t) = \int d^dq q^2iD_K(q,t,t)  &= \int d^d q q^2 \frac{1-e^{-2\lambda_q t}}{2\lambda_q} =  1- \left(\frac{r}{T}\right)^{d/2}\Gamma(-d/2,2rt)\label{J2},
\end{align}
where $\Gamma(-d/2, x)$ is the incomplete Gamma function defined by,
\begin{equation}
\Gamma(-d/2,x) = \int_x^\infty  dy y^{-d/2 - 1} e^{-y}.
\end{equation}
\begin{align}
B(t,s)=\int d^dq q^2
	\biggl[
		\frac{
			e^{-2\lambda_q(t-s)}-e^{-2\lambda_q t}
			}{
			2q^2
		}
	\biggr] =
	\frac{e^{-2r(t-s)}}{(T(t-s))^{d/2+1}}
	 - \frac{e^{-2rt}}{(Tt)^{d/2+1} },
\label{J3}
\end{align}
and
\begin{align}
C(t,s) &= \int_0^s dt' \biggr[ 	\frac{e^{-2r(t-s)}}{T(t-s)^{d/2+1}}
	 - \frac{e^{-2r(t)}}{T(t)^{d/2+1} }\biggr] \nonumber\\
	 &= \frac{1}{T}\left(\frac{2r}{T}\right)^{d/2}
	\left[	
		\Gamma\left(-\frac{d}{2}, 2r(t-s) \right)
		-
		\Gamma\left(-\frac{d}{2}, 2rt \right)
	\right] - \frac{s e^{-2rt}}{(Tt)^{d/2+1}}.
	\label{J4}
\end{align}
In the limit $ r\rightarrow 0$ these formulae become.
\begin{align}
A(t) &= 1- \biggl(\frac{1}{t+1}\biggr)^{d/2},\\
B(t,s)&= \frac{1}{(t-s+1)^{1+d/2}}-\frac{1}{(t+1)^{1+d/2}},\\
C(t,s)&= \frac{2}{d}\biggl[\frac{1}{(t-s+1)^{d/2}}-\frac{\left(1+ \frac{d}{2}\frac{s}{t+1}\right)}{(t+1)^{d/2}}\biggr].
\end{align}

\section{Time-Dependent detuning} \label{tdq}
We consider now the Hamiltonian as in Eq.~\eqref{Hf}, but where $U$ is an arbitrary function of time
\begin{equation}
H= H_0 + \frac{U(t)}{N}\sum_q \Delta^\dagger_q \Delta_q.
\end{equation}

The analysis then precedes identically as with the above except that the equation of motion for $D_R$ is modified to,
\begin{equation}
\int_{t''}^t\!dt'\, \left[U(t)\delta(t,t') - \Pi_R(t,t')\right] D_R(t',t'') =  \delta(t-t'').
\end{equation}
Making the same approximations as in time independent case leads to
\begin{gather}
\left[\partial_t +\lambda_q(t)\right]D_R(t,t') = Z\delta(t-t'),\\
D_R(t,t') = Z\exp\left(-\int_{t'}^{t}ds \lambda_q(s)\right),\\
\lambda_q(t) = r(t) + \frac{q^2}{2M};\qquad r(t) \equiv U(t) - \tilde{\Pi}_R(q=0,\omega = 0).
\end{gather}
The expression for $D_K$ then follows from $D_R\circ \Pi_K \circ D_A$. Using the approximation $\Pi_K(t-t') \sim iTZ^{-1}\delta(t-t')$ as before we obtain
\begin{gather}
\begin{aligned}
iD_K(t,t') &= TZ^{-1}\int_0^{ {\rm min}(t,t')} \!ds \, D_R(t,s) D_A(s,t')\\
		&= TZ \int_0^{ {\rm min}(t,t')} \!ds \exp\left(-\int_s^t du \lambda_q(u) - \int_s^{t'} du \lambda_q(u)\right)\\
		&= \left[D_R(t,t')F_q(t') + F_q(t)D_A(t,t')\right],
\end{aligned}\\
	F_q(t)\equiv 2T\int_0^t\!ds\, \exp\left( -2 \int_s^t \!du\,\lambda_q(u)\right).
\end{gather}
Thus the functions $D_K$ and $D_R$ have the same structure as in the time independent case, and the previous manipulations may be repeated with the substitution: $\exp\left(-\lambda_q(t-t')\right)\rightarrow \exp\left(-\int^t_{t'}ds \lambda_q(s)\right)$.

The end result is a generalization of Eq.~\eqref{kinfin}:
\begin{equation}
\partial_t J^i - E^i(t) = \sum_q\left\{-q^2 F_q(t)
+ \alpha q^2 \lambda_q(t)\int_0^t\!ds\,
\left[
	J^i(s) +\frac{1}{2}\int_s^t dt' E^i(t')
\right]
e^{-2\int_s^t\!du\, \lambda_q(u)} F_q(s)
\right\}
\end{equation}

This is of the same form as Eq.~\eqref{J1} but with the redefinitions
\begin{align}
A(t) &= \sum_q q^2 F(q) \\
	&\propto \int d^d q q^2 T\int_0^t\!ds\, \exp\left( -2 \int_s^t \!du\,\lambda_q(u)\right)\\
	&\propto \int d(q^2) q^{d} T\int_0^t \!ds \,
	\exp\left(
	-\frac{q^2}{M}(t-s) -2 \int_s^t \!du\, r(u)
	\right) \\
	&\propto \int_0^t\!ds\, \frac{e^{
	-2 \int_s^t \!du\, r(u)
	}}{(t-s)^{d/2+1} },
\end{align}

\begin{align}
B(t,s) &= \sum_q\left\{ q^2\lambda_q e^{-2\int_s^t\!du\, \lambda_q(u)} F_q(s)\right\}\\
	 &\propto \int d^d q
	 \left\{
	 	q^2
	 	\left(
	 		\frac{q^2}{2M} + r(t)
	 	\right)
		e^{-2\int_s^t\!du\,\lambda_q(u)}
		\left[
			T\int_0^s dt' e^{-2\int_{t'}^s \! du \, \lambda_q(u)	}
		\right]	
	 \right\}\\
	&\propto
	\int d(q^2) q^{d}
	 \left\{
	 	\left(
	 		\frac{q^2}{2M} + r(t)
	 	\right)
		\int_0^s dt' e^{-2\int_{t'}^t \! du \, \lambda_q(u)	}
	 \right\}\\
	&\propto
	\int d(q^2) q^{d}
	 \left\{
	 	\left(
	 		\frac{q^2}{2M} + r(t)
	 	\right)
		\int_0^s dt' e^{-\frac{q^2}{M}(t-t') -2\int_{t'}^t \! du \, r(u)	}
	 \right\}\\
	 &\propto
	 \int_0^s\!dt'\,	
	 \left(
		 \frac{1+d/2}{(t-t')^{d/2+2} }  +
		 \frac{2r(t)}{(t-t')^{d/2+1} }
	 \right)
	 e^{-2\int_{t'}^{t}\!du\, r(u)}
	 \\
	 &=
	 \int_0^s\!dt'\,	
	 \left(
		 \frac{\left(1+\frac{d}{2}\right) + 2 r(t)(t-t')}{(t-t')^{d/2+2} }
	 \right)
	 e^{-2\int_{t'}^{t}\!du\, r(u)},
\end{align}
and $C(t,s) = \int_0^s ds' B(t,s')$.

\subsubsection{Steady state behavior}

Let us consider steady state in the presence of a fixed detuning. Here,
the conductivity $\sigma(\omega,t)$ becomes time-independent with a peak at zero frequency, the
height of which may be estimated by taking the limit $r t \gg 1$ and neglecting the $B$ term, giving,
\begin{equation}
\left(\partial_t + \frac{1}{\tau_r} \right) J = \frac{\alpha}{2\tau_r}\int_0^\infty  dt' C(t,t')E(t').
\end{equation}
Since the integral on the RHS varies slowly with time, then in the DC limit this gives,
$J = (\alpha/2) \int dt' C(t,t') E(t')$. Therefore the conductivity has a tail of the form
$\sigma \left(t +\tau, t\right) \sim \alpha C(\tau)$.  Now $C(\tau)$ decays as $1/\tau$ until $\tau\sim r^{-1}$.
Thus $\sigma(\omega =0 )\sim \int d\tau C(\tau) \sim\alpha \log{(r/T)}$. 
At $r=0$, the apparent logarithmic divergence is cut-off by terms higher order in $1/N$.

\end{document}